\documentclass[aps,amssymb,11pt,eqsecnum]{revtex4}

\usepackage{graphicx,graphics}
\usepackage{enumerate}
\usepackage{subfigure}
\usepackage{amsmath}
\usepackage{amsfonts}
\usepackage{amssymb}
\usepackage{amsthm}
\usepackage{psfrag}

\catcode`\@=11
\def\numberbysection{\@addtoreset{equation}{section}
         \def\theequation{\arabic{section}.\arabic{equation}}}
\numberbysection

\def\be{\begin{equation}}
\def\ee{\end{equation}}
\def\bea{\begin{eqnarray}}
\def\eea{\end{eqnarray}}

\begin{document}

\title{The problem of predecessors on spanning trees}

\author{V.S. Poghosyan$^{1}$, V.B. Priezzhev$^2$ }
\affiliation{
$^1$Institut de Physique Th\'{e}orique, Universit\'{e} catholique de Louvain, B-1348 Louvain-La-Neuve, Belgium,\\
$^2$Bogoliubov Laboratory of Theoretical Physics,
Joint Institute for Nuclear Research, 141980 Dubna, Russia
}

\begin{abstract}
We consider the equiprobable distribution of spanning trees on the square lattice. All bonds of each tree can be oriented uniquely
with respect to an arbitrary chosen site called the root. The problem of predecessors is finding the probability that a path along
the oriented bonds passes sequentially fixed sites $i$ and $j$. The conformal field theory for the Potts model predicts the fractal
dimension of the path to be 5/4. Using this result, we show that the probability in the predecessors problem for two sites separated
by large distance $r$ decreases as  $P(r) \sim r^{-3/4}$. If sites $i$ and $j$ are nearest neighbors on the square lattice, the
probability $P(1)=5/16$ can be found from the analytical theory developed for the sandpile model. The known equivalence between the
loop erased random walk (LERW) and the directed path on the spanning tree says that $P(1)$ is the probability for the LERW started
at $i$ to reach the neighboring site $j$. By analogy with the self-avoiding walk, $P(1)$ can be called the return probability.
Extensive Monte-Carlo simulations confirm the theoretical predictions.
\end{abstract}

\maketitle
\noindent \emph{E-mail}: vahagn.poghosyan@uclouvain.be, priezzvb@theor.jinr.ru

\section{Introduction and Main Results}

In the graph theory, the spanning tree of connected graph $G$ is a connected subgraph of $G$ containing all vertices of $G$ and having no cycles.
Numerous applications of spanning trees began with the seminal Kirchhoff's problem solved in 1847 and then spread out many branches of
mathematics and theoretical physics. In the statistical mechanics, spanning trees are related to the Potts model \cite{Wu},
the dimer model \cite{FishSt}, the sandpile model \cite{SOC} and many others.
A relation between lattice models of statistical mechanics and spanning trees via the Tutte polynomial has been established by Fortuin and Kasteleyn \cite{FortKast}.

The Kirchhoff theorem claims that the number of spanning trees of connected graph $G$ is a cofactor of the Laplacian  matrix $\Delta$ of graph $G$.
If one deletes any row and any column from $\Delta$, one obtains a matrix $\Delta^*$ which gives the number of spanning trees as $\det \Delta^*$. The
determinantal structure allows easy calculation of local characteristics of the spanning trees, for instance, the average number of vertices
with given number of adjacent bonds. A characterization of non-local objects in the spanning tree is not so simple. One of such the objects
is the chemical path defined as a path along two or more bonds of the tree. The fractal dimension of a long chemical path on the two-dimensional
lattice has been calculated by means of a mapping of the spanning tree configurations onto the Coulomb gas model.

A closely related object is the loop erased random walk (LERW) on the two-dimensional lattice \cite{LawlerLERW} which was proven to be equivalent to the directed
chemical path of the spanning tree of the same lattice \cite{MajLERW,KenyonLERW}. In this paper, we consider a problem arising in the theory of LERW  and equally distributed
spanning trees: given two lattice sites $i$ and $j$, what is the probability that the LERW or the directed chemical path passes  $i$ and $j$.
If site $i$ is passed first, we say that $i$ is the predecessor of $j$ and coin the mentioned problem as the predecessor problem. Surprisingly,
the problem has no exact solution in the general case. Only two limiting cases are available: (a) If sites $i$ and $j$ are separated by
large distance $r$ , the asymptotics of $P(r)$ can be found from known results on the fractal dimension of the chemical path; (b) If points
$i$ and $j$ are nearest neighbors of the square lattice, the seeking probability can be found from the theory of sandpiles \cite{Priez} (see also \cite{jpr}).

The asymptotic behavior of $P(r)$ for large distance $r$ follows directly from the definition of fractal dimension.
Indeed, consider a large square lattice $\mathcal{L}$ and the set of uniformly distributed spanning trees on $\mathcal{L}$.
We assume that the root is situated at the boundary of $\mathcal{L}$.
Consider site $i$ in the bulk of the lattice and some circle contour $C$ of radius $R$ with the center in $i$. Let $\Pi$ be a directed
chemical path from $i$ to the root along the oriented bonds of a tree.
All points of the subset of $\Pi$ inside $C$ are descendants of $i$.
In accordance with the definition of the fractal dimension of the directed path on the spanning tree, the number of the descendants inside
$C$ is proportional to $R^{5/4}$ (see Majumdar \cite{MajLERW}).
The probability that point $i$ is the predecessor of point $j$ lying at distance $r$ from $i$ is the density of descendants $\rho(r)$.
Thus, we have

\begin{equation}
\label{integral}
\int_1^R \rho(r)rdr\sim R^{5/4}
\end{equation}
from where we conclude that $\rho(R)\sim R^{-3/4}$.

As it was mentioned before, the problem of predecessors for an arbitrary disposition of two lattice points is not solved.
In Section 2 we concentrate on a particular problem of probability $P(1)$ when points $i$ and $j$ are nearest neighbors of the square lattice.
An essential element of the theory of sandpiles is the probability distribution of sites having 0, 1, 2 and 3 predecessors among the nearest neighbors.
The corresponding probabilities are denoted by $X_0,\, X_1,\, X_2$ and $X_3$.
Having explicit expressions for these values, we obtain $P(1)$ as their combination and get an unexpectedly simple result $P(1)=5/16$.
In Section 3 we relate this result to the return probability of the LERW. Section 4 contains results of the Monte-Carlo verifications.

\section{The problem of predecessors for nearest neighbors}

The spanning tree enumeration method, namely, the Kirchhoff theorem is proved to be a powerful mathematical instrument for the investigation of
various combinatorial problems of the theoretical physics.
In the last decade, it has been developed and adapted for the calculation of height probabilities of the Abelian sandpile model \cite{majdhar1,Priez,jpr}.
The Abelian sandpile model is a stochastic dynamical system, which describes the phenomenon of self-organized criticality.
During the evolution the system falls into a subset of all possible states, called the subset of recurrent states.
The problem is to calculate analytically various observable values in the recurrent state, such as height probabilities $P_i,\, i=1,2,3,4$ at a fixed site and
height correlations between distinct fixed sites \cite{correlations}.

It was shown that the calculation of height probabilities in the Abelian sandpile can be reduced to the calculation of $X_0,\, X_1,\, X_2$ and $X_3$
in the spanning tree model.
The exact relation between these quantities is given by
\begin{equation}
P_1 =       \frac{X_0}{4};\quad
P_2 = P_1 + \frac{X_1}{3};\quad
P_3 = P_2 + \frac{X_2}{2};\quad
P_4 = P_3 + X_3.
\label{siteprob}
\end{equation}

Majumdar and Dhar \cite{majdhar1} have found in 1991 the probability of height 1,
constructing the corresponding defect lattice for the situation when a site $i_0$ has no predecessors and calculating the determinant
of the defect matrix $\Delta ^*$.
A technique for computing the numbers $X_1, X_2, X_3$ has been devised in \cite{Priez}.
The results are (see also \cite{jpr} for details)
\begin{equation}
X_0 = \frac{8 (\pi -2)}{\pi ^3};\quad
X_1 = \frac{3}{4}+\frac{48}{\pi ^3}-\frac{15}{\pi ^2}-\frac{3}{2 \pi };\quad
X_2 = \frac{1}{4}-\frac{48}{\pi ^3}+\frac{6}{\pi ^2}+\frac{3}{\pi };\quad
X_3 = \frac{16}{\pi ^3}+\frac{1}{\pi ^2}-\frac{3}{2 \pi },
\end{equation}
which give
\begin{equation}
P_1 = \frac{2 (\pi -2)}{\pi ^3};\quad
P_2 = \frac{1}{4}+\frac{12}{\pi ^3}-\frac{3}{\pi ^2}-\frac{1}{2 \pi };\quad
P_3 = \frac{3}{8}-\frac{12}{\pi ^3}+\frac{1}{\pi };\quad
P_4 = \frac{3}{8}+\frac{4}{\pi ^3}+\frac{1}{\pi ^2}-\frac{1}{2 \pi }.
\end{equation}

\begin{figure}[h!]
\includegraphics[width=100mm]{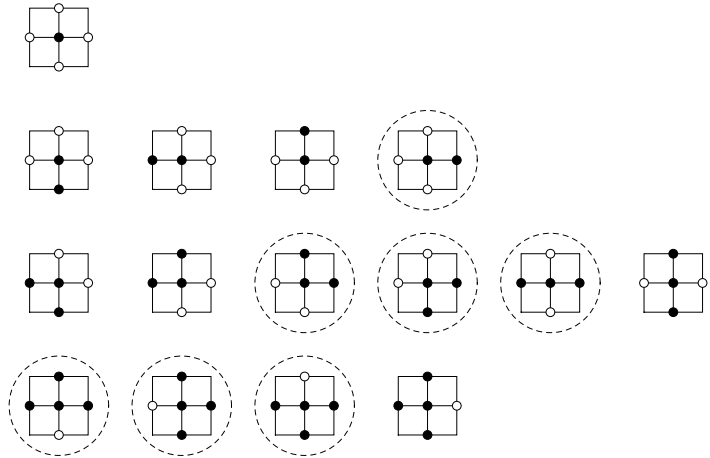}
\caption{\label{fig1} All possible situations with a fixed vertex $i_0$ (the central vertex in diagrams)
to have various nearest neighbouring predecessors on square lattice.
By black color we denote vertices, which are predecessors of $i_0$.
White means that the corresponding vertex is not a predecessor of $i_0$.
On the $k+1$st row ($k=0,1,2,3$) situations with $k$ predecessors are presented.
By the dashed line are circled the configurations for which the right neighboring vertex is a predecessor of $i_0$.
}
\end{figure}

Now consider the problem of predecessors for nearest neighboring sites. First we fix a site $i_0$ in the bulk of the square lattice.
Denote its right nearest neighboring site by $j_0$.
Next consider four various cases, when $i_0$ has exactly $k$ nearest neighboring predecessors $(k=0,1,2,3)$ (see Fig \ref{fig1}).
For $k=0$ the site $j_0$ trivially is not a predecessor of $i_0$.
For $k=1$, we have $1$ of $4$  equivalent situations when $j_0$ is the predecessor of $i_0$. For $k=3$, we have $3$ of $4$
equivalent situations when $j_0$ is the predecessor.
The crucial case is the $k=2$. Here we have $6$ situations, but not all of them are equivalent.
On the other hand, we can select two groups of $4$ and $2$ elements (the first four and the last two in the third line of Fig. 1)
so that elements in each group  are equivalent.
We are looking for the situations where $j_0$ is a predecessor of $i_0$.
There are $2$ encircled elements from the first group and one from the second group, which correspond to the desired situations.
Thus, if we take the linear combination of $X_1,\, X_2$ and $X_3$ with coefficients $1/4$, $1/2$ and $3/4$ correspondingly,
we get the desired probability $P(1)$ that $j_0$ is the predecessor of $i_0$:
\begin{equation}
P(1) = \frac{1}{4} X_1 + \frac{1}{2} X_2 + \frac{3}{4} X_3 = \frac{5}{16}.
\end{equation}


\section{Return probability for the loop erased random walk}
\label{sec2}

Consider a finite square lattice $\mathcal{L}$ with vertex set $V$ and edge set $E$.
Given $\mathcal{P}=[u_0, u_1, u_2, \ldots, u_n]$ a path in $\mathcal{L}$,
its loop-erasure $\mathrm{LE}(\mathcal{P}) = [\gamma_0, \gamma_1, \gamma_2, \ldots, \gamma_m]$ is defined by chronologically removing loops from $\mathcal{P}$.
Formally, this definition is as follows.
We first set $\gamma_0 = u_0$. Assuming $\gamma_0,\ldots,\gamma_k$ have been defined, we let $s_k = 1 + \mathrm{max} \{j: u_j = \gamma_k\}$.
If $s_k = n+1$, we stop and $\mathrm{LE}(\mathcal{P}) = [\gamma_0, \gamma_1, \gamma_2, \ldots, \gamma_m]$ with $m = k$.
Otherwise, we let $\gamma_{k+1} = u_{s_k}$.
Note that the order in which we remove loops does matter, and it follows from the definition, we remove loops as they are created, following the path.
A walk, obtained after applying the loop-erasure to a simple random walk path is called Loop-Erased Random Walk (LERW).
Since the infinite simple random walk on finite connected undirected graphs is recurrent, the infinite LERW is not defined.
On the other hand, we can fix a subset $W \subset V$ of vertices and define LERW from a fixed vertex $u_0$ to $W$.
To do that we take a path of a simple random walk started at $u_0$ and stopped upon hitting $W$, after that we apply loop-erasure.

Wilson  established an algorithm to generate uniform spanning trees, which uses LERW \cite{Wilson}.
It turns out to be extremely useful not only as a simulation tool, but also for theoretical analysis.
It runs as follows.
Pick an arbitrary ordering $V = \{ v_0, v_1, \ldots, v_N \}$ for the vertices in $\mathcal{L}$.
Let $S_0={v_0}$.
Inductively, for $i = 1,2,\ldots,N$ define a graph $S_i$ to be the union of $S_{i-1}$ and a (conditionally independent) LERW path from $v_i$ to $S_{i-1}$.
Note, if $v_i \in S_{i-1}$, then $S_i=S_{i-1}$. Then, regardless of the chosen order of the vertices, $S_N$ is a UST on $\mathcal{L}$ with root $v_0$.
If we take as an initial condition $S_0 = W$, with some $W \subset V$, then the generated structures will be spanning forests with set of roots $W$.
The spanning forest with fixed set of roots can be considered as a spanning tree, if we add an auxiliary vertex and join it to all the roots.

Now consider the Wilson algorithm on  $\mathcal{L}$ with the set of boundary vertices
$\partial\mathcal{L}$ and take $S_0 = \partial\mathcal{L}$.
When the size of the lattice tends to infinity, the boundary effects will vanish, so we can neglect the details of the boundary.
So we will not distinguish between spanning forests and spanning trees.
It follows from the Wilson algorithm for a fixed site $i_0$, that if we start a LERW from $i_0$ upon hitting the boundary $\partial\mathcal{L}$,
we will generate a path $\ell$ of a spanning tree from $i_0$ to the boundary (see also \cite{MajLERW},  \cite{lswLERW}).
All vertices on the path $\ell$ form the set of descendants of $i_0$.
So, if a fixed vertex $j_0$ belongs $\ell$, $i_0$ is a predecessor of $j_0$.

Consequently, the probability $P(j_0-i_0)$ that $i_0$ is a predecessor of $j_0$ in a randomly taken (from the uniform distribution)
spanning tree on the large square lattice equals to the probability that the LERW started from $i_0$ passes $j_0$. In the particular case
$|j_0-i_0|=1$, the probability $P(1)=5/16$ calculated in the previous section is the return probability for the LERW.

\section{Monte-Carlo simulations}

\begin{figure}[h!]
\includegraphics[width=80mm]{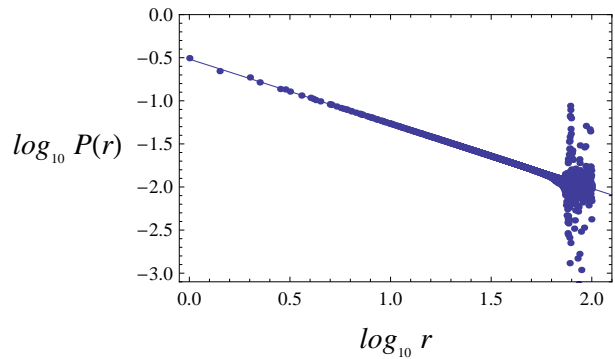}
\caption{\label{fig2} The results of Monte-Carlo simulations for the probability $P(r)$.
}
\end{figure}

Consider finite $2 N + 1 \times 2 N + 1$ square lattice $\mathcal{L}_N$.
Denote its central vertex by $i_0$ and assume that it is an origin of the coordinate system.
We deliberately took an odd-odd lattice to provide the symmetry which guarantee the most efficient vanishing of boundary effects for large lattices.
After generating a large amount of LERWs starting from $i_0$, we get an approximation of the function $P_N(j_0-i_0) \equiv P_N(j_0)$.
Given fixed $j_0$, we can extrapolate $P_N(j_0)$, tending $N$ to infinity and get asymptotical function $P(j_0)$.
Assume that the Euclidean distance between the origin $i_0$  and $j_0$ is $r$ and coordinates of $j_0$ are  $(r \cos \varphi,\, r \sin \varphi)$.
The Monte-Carlo simulations and Coulomb gas arguments show that the asymptotical behaviour of the function $P(j_0)$ for large $r$ $(r \gg 1)$
does not depend on $\varphi$.
So, for $r\gg 1$ we have $P(j_0) \simeq P(r)$.
Fig. \ref{fig2} shows the behaviour of $P(r)$ for various $j_0$ on the log-log scale, obtained from Monte-Carlo simulations.
From this result we conclude that $P(r)$ decreases as a power function:
\begin{equation}
P(r) \simeq \frac{C}{r^{\alpha}},
\end{equation}
with $\alpha \approx 0.751$ and $C \approx 0.305$. During the simulations, we took sizes up to $N = 100$ and number of simulations $10^8$.
The obtained results are in agreement with $\alpha = 3/4$ which follows from the Coulomb gas arguments mentioned above.
The effective Monte-Carlo algorithm allows evaluation of probabilities $P(j_0-i_0)$ for arbitrary finite $j_0,i_0$. At the same time, the
analytical calculation of $P(r)$ for $r>1$ remains a difficult unsolved problem.

\section*{Acknowledgments}
This work was supported by the Russian RFBR grant No 09-01-00271a, and by the Belgian Interuniversity Attraction Poles Program P6/02,
through the network NOSY (Nonlinear systems, stochastic processes and statistical mechanics).
We would like to thank P. Ruelle for helpful discussions.
The Monte-Carlo simulations were performed on Armenian Cluster for High Performance Computation (ArmCluster, www.cluster.am).

\end{document}